\documentclass[twocolumn]{pasj00}
\usepackage{graphicx}
\SetRunningHead{}{}
\Received{2005/11/22}
\Accepted{2006/05/01}
\begin{document}

\title{Dynamical friction on satellite galaxies}
\author{Michiko \textsc{Fujii}}
\affil{Department of Astronomy, Graduate School of Science,
  the University of Tokyo, Tokyo, 113}
\email{fujii@astron.s.u-tokyo.ac.jp}

\author{Yoko \textsc{Funato}}
\affil{General Systems Studies, Graduate Division of International and
  Interdisciplinary Studies,\\ University of Tokyo, Tokyo, 153}
\email{funato@chianti.c.u-tokyo.ac.jp}

\and
\author{Junichiro \textsc{Makino}}
\affil{Department of Astronomy, Graduate School of Science,
  the University of Tokyo, Tokyo, 113}
\email{makino@astron.s.u-tokyo.ac.jp}

\maketitle

\begin{abstract}
For a rigid model satellite, Chandrasekhar's dynamical friction
formula describes the orbital evolution quite accurately, when the
Coulomb logarithm is chosen appropriately. However, it is not known if
the orbital evolution of a real satellite with the internal degree of
freedom can be described by the dynamical friction formula. We
performed $N$-body simulation of the orbital evolution of a
self-consistent satellite galaxy within a self-consistent parent
galaxy. We found that the orbital decay of the simulated satellite is
significantly faster than the estimate from the dynamical friction
formula. The main cause of this discrepancy is that the stars stripped
out of the satellite are still close to the satellite, and increase
the drag force on the satellite through two mechanisms.  One is the
direct drag force from particles in the trailing tidal arm, a
non-axisymmetric force that slows the satellite down.  The other is
the indirect effect that is caused by the particles remaining close to the
satellite after escape. The force from them enhances the wake caused in
 the parent galaxy by dynamical
friction, and this larger wake in turn slows the satellite down more
than expected from the contribution of its bound mass.  We found these
two have comparable effects, and the combined effect can be as large
as 20\% of the total drag force on the satellite.
\end{abstract}

\section{INTRODUCTION}

The evolution of satellite galaxies has been studied by a number of
researchers both theoretically and using numerical
simulations. However, even though it is  a basic and  simple
problem, our understanding is still rather limited.

Using $N$-body simulation of a rigid satellite within an $N$-body
model of the parent galaxy, \citet{B99} found that the
orbital eccentricity of a satellite galaxy tends to be roughly
constant. Previous theoretical studies based on Chandrasekhar's
dynamical friction formula \citep{C43} predicted
circularization of the orbit. Thus, there was rather serious
qualitative difference between the simulation result and the
theoretical model.

In $N$-body calculations of \citet{B99}, the parent galaxy was modeled
as an $N$-body system, while the satellite was treated as one massive
softened particle. Thus, the tidal mass loss was ignored in their
calculation. \citet{JB00} performed a self-consistent $N$-body
simulation of the evolution of a satellite galaxy, in which both the
satellite and the parent galaxy were treated as $N$-body system. They
compared the result with that of a semianalytical model, in which the
orbit of the satellite evolved through the dynamical friction
expressed by Chandrasekhar's formula. The agreement between the
simulation and semianalytic model was not
good. \citet{Velazquez1999} performed similar comparison, and found
that it was possible to make simulation result and semianalytic model
agree to each other, if they use the Coulomb logarithm as a fitting
parameter. \citet{TB2001} constructed a more sophisticated model for
the evolution of the satellite, and demonstrated that it could
reproduce the simulation results of \citet{Velazquez1999} quite
accurately.

The dynamical friction formula is given by
\begin{eqnarray}
\frac{d{\mbox{\boldmath $v$}}_{\rm s}}{dt}= -16\pi^2 G^2 m(M_{{\rm
    s}}+ m) \log \Lambda 
\frac{{\int_0^{v_{\rm s}}} f(v_{\rm m})
v_{\rm m}^2d{v_{\rm m}}}
{|{\mbox{\boldmath $v$}}_{\rm s}|^3} {\mbox{\boldmath $v$}}_{\rm s}.
\label{eq:1}
\end{eqnarray}
Here, $v_{\rm s}$ is the velocity of the satellite, 
$G$ is the gravitational constant, $M_{\rm s}$ and $m$ are  the masses of the satellite galaxy
and field particles of the parent galaxy,
and $f(v)$ is the distribution function of field
particles at the position of the satellite. We assumed that the
velocity distribution is isotropic, which is true at least for the
initial model of the parent galaxy we consider in this paper.  
The term  $\log \Lambda$ is the Coulomb logarithm 
given by
\begin{equation}
\log \Lambda = \log \left( \frac{b_{\rm max}}{b_{\rm min}} \right).
\label{eq:logLambda}
\end{equation}
Here, $b_{\rm max}$ and $b_{\rm min}$ are the maximum and the minimum 
impact parameters for gravitational encounters between the satellite
galaxy and the field stars in the parent galaxy. For the lower cutoff,
it is natural to set $b_{\rm min}$ to the order of the virial radius of
the satellite galaxy (White 1976). For the upper cutoff, in many studies
the size of the parent galaxy $R$ has been used (e.g., Murai \& Fujimoto
1980; Helmi et al. 1999; Johnston et al. 1995). \citet{JB00} followed
this tradition and used $\log \Lambda = 8.5$, while
\citet{Velazquez1999} varied it to obtain the best agreement, and the
value they used was $1 \sim 2$. Clearly, however, one cannot use the
Coulomb logarithm as a fitting parameter, since its value should be
determined from the mass distributions of the satellite and the parent
galaxy. 

Hashimoto et al. (2003, hereafter H03) showed that
Chandrasekhar's dynamical friction formula does give the result which is
consistent with the simulation result, when the integration over the
impact parameter is carried out
correctly. Equation (\ref{eq:logLambda}) was obtained assuming that
the distribution of the field stars is uniform and infinite. We need to
apply the upper cutoff since the integration would
diverge without the cutoff. However, the actual $N$-body system has a
finite size so that the integration over the mass distribution could not
diverge. They pointed out that, if this integration over the mass
distribution of the parent galaxy is correctly performed, the
discrepancy between $N$-body simulation result and the model with
analytic estimate disappears. The exact integration over all
encounters generally resulted in the Coulomb logarithm smaller than
the value given by equation (\ref{eq:logLambda}) with $b_{\rm max}=R$,
simply because the density drops off at large radii.  The use of
equation (\ref{eq:logLambda}) with $b_{\rm max}=R$
is the same as  to assume that the parent
galaxy is a sphere of radius $R$ with a constant density same as the
local density around the satellite. Also, it implies that the orbit of
the satellite is a straight line. However, since the density drops off
and the orbit of the satellite is not a straight line, the actual
dynamical friction is smaller. If we set the cutoff radius as the
distance of the satellite from the center of the parent galaxy,
equation (\ref{eq:1}) gives the result in good agreement with the
$N$-body simulation.

H03 resolved the discrepancy between the theoretical and numerical
results for the case of a rigid satellite. The formula they proposed,
and a modified version of it \citep{Z03} are used in recent studies of
the evolution of subhalos and satellite galaxies. In such studies, a
self consistent $N$-body model is used for the satellite (or subhalo)
and its orbit is integrated numerically using the dynamical friction
formula, to express the time variation of the tidal field.  This
technique has been used by a number of researchers (Bullock \&
Johnston 2005; Johnston et al. 1995; Ibata \& Lewis 1998; Portegies
Zwart et al. 2004).

However, there is no guarantee that the orbital evolution of a live
satellite is correctly described by the dynamical friction formula of
H03. The results of previous studies \citep{JB00,Velazquez1999} suggest
that the disagreement between simulation and semianalytic
model with dynamical friction formula is due to the incorrect choice
of the Coulomb logarithm. Even so, there has been no study in which
the dynamical friction formula of H03 was directly compared to a fully
self-consistent $N$-body simulation. That means the results can
be very different.

In this paper, we performed such a direct comparison between fully
self-consistent $N$-body simulation and semianalytic model. We found
a serious disagreement, and we investigated the reason of that
disagreement. Our main findings can be summarized as follows. The
disagreement does exist, but in the direction opposite to that
observed in previous studies. The orbital decay in the $N$-body
simulation was significantly faster than that in the semianalytic
model with the dynamical friction formula of H03.  We investigated the
cause of this disagreement, and found that the main causes are the
effect of particles which are stripped from the main body of the
satellite by the tidal field of the parent galaxy. These escaped
particles exert drag forces to the main body of the satellite in two
different ways. The first one is the direct gravitational force. Since
the distribution of the escaped particles is not axisymmetric around
the center of the parent galaxy, their gravity can change the orbital
energy of the satellite, and it turned out that the forces integrated
over all escaped particles effectively works as a drag force. The
second one is the enhancement of the dynamical friction. Escaped
particles typically go away from the satellite rather slowly. In other
words, they move together with the satellite for a rather long
time. Thus, from the point of view of the field particles, these
``escaped'' particles are still effectively part of the satellite,
since the gravitational force they feel is the combined effect of the
main body of the satellite and ``escaped'' particles which still are
close to it. The dynamical friction on the body of the satellite is
therefore bigger than what would be there if there are no such escaped
particles. We found that these two effects have comparable strength
and the combined effect explains the disagreement between the $N$-body
simulation and semianalytic calculation.

This paper is organized as follows. In
section 2 we describe the simulation method and initial conditions. In
section 3 we give the result. The semianalytic model gave the orbital
decay significantly slower than that 
observed in the $N$-body simulation.  In section 4 we discuss the
cause of this discrepancy. Section 5 is for a summary and discussions.

\section{Numerical methods}

\subsection{Initial condition}

We consider a simple problem of one spherical satellite galaxy
orbiting in a spherical parent galaxy. This is essentially the same
problem as that studied by H03. In the following we describe the initial
model. 

We adopted a King model with non-dimensional central potential
$W_0=9$ as the model
of the parent galactic halo and $W_0=7$ as that of the
satellite halo. The system of units is the Heggie unit (Heggie \&
Mathieu 1986), where the gravitational constant $G$ is 1 and the mass
and the binding energy of the parent galaxy are 1 and 0.25, respectively.
Initially, the satellite is placed at distance 1.5 from the center of
the parent galaxy, with the velocity of 0.45. Assuming that the parent
galaxy represents our Galaxy with total mass $M=10^{12}M_{\odot}$ and
the circular velocity $V_{\rm c}=250$ km s$^{-1}$, the initial distance
and velocity of the satellite galaxy are 60 kpc and 140 km s$^{-1}$.
Unit time in the Heggie unit corresponds to 130 Myr.

In table 1, we summarize the model parameters and
initial conditions of our $N$-body simulations. Most of the parameters
are the same as those used in H03. We chose the
initial velocity slightly larger than what is used in H03, to keep
the mass loss rate smaller. This choice allowed us to follow the
evolution of satellite for more than 10 orbits.

\subsection{$N$-Body Simulation}

In the $N$-body simulation, both the parent galaxy and the satellite
were expressed as self-consistent $N$-body models. 
The number of particles $N$ of the parent is $10^6$ and that of the
satellite is $5 \times 10^4$. The number of particles in the satellite
should be large enough that the relaxation effect does not seriously
affect the mass loss from the satellite. Since the initial half-mass
relaxation time of the satellite is about 160 in our system of units,
relaxation effect is small.

The number of particles in the parent galaxy should be determined so
that the two-body relaxation effect on particles in the parent galaxies
and that in satellite galaxies are small compared to the velocity
dispersion of particles. Since
the velocity dispersion of particles in the parent galaxy is much
higher than the internal velocity dispersion of satellite particles,
we only need to consider the heating of satellite particles due to
encounters with particles in the parent galaxy. The timescale of this
heating, $T_h$, is expressed as 

\begin{equation}
\label{eq:heating}
T_h = t_{rh,p} \frac{\sigma_s^2}{\sigma_p^2},
\end{equation}
for the first-order approximation, 
where $t_{rh,p}$ is the half-mass
relaxation timescale of the parent galaxy, $\sigma_s$ and $\sigma_p$
are the velocity dispersions of the satellite and the parent galaxy,
respectively. For our choice of initial model and number of particles,
$T_h \sim 10^3$ and it is sufficiently longer than the duration of the
simulation.

We used a Barnes-Hut treecode (Barnes \& Hut 1986; Makino 2004) on
GRAPE-6A (Fukushige et al. 2005). We used opening angle $\theta=0.75$
with center-of-mass (dipole-accurate) approximation. The maximum group
size for GRAPE calculation (Makino 1991) is 8192. For the time
integration a leapfrog integrator with a fixed stepsize of $\Delta
t=1/256$ is used. The potential is softened using the usual Plummer
softening, with the softening length $\epsilon=0.00625$. This same
softening is used for all interactions.  The total energy was conserved
to be lower than 0.02 \% throughout the simulation.

To calculate the mass and orbit of the satellite, we need to identify
the particles which belong to the satellite. We determine these
particles by an iterative procedure (Funato et al. 1993). One particle
belongs to the satellite if its binding energy to the satellite is
negative. Potential energy is calculated using all other particles
which belong to the satellite, and kinetic energy is calculated
relative to the center-of-mass motion of the satellite.

\begin{table}
\begin{center}
\caption{Model Parameters of $N$-body Simulations}  
\begin{tabular}{lcc}
\hline \hline
Parameters & Parent & Satellite \\ \hline
Galactic halo & King 9 & King 7 \\ 
Total mass & 1.0 & 0.01  \\ 
Binding energy & 0.25 & $0.25 \times 10^{-3}$ \\ 
Half-mass radius & 0.98 & 0.081 \\
N    & $10^6$& $5\times 10^4$ \\ \hline
Initial position & & ( 1.5, 0, 0) \\ 
Initial velocity & &( 0, 0.45, 0) \\ \hline
\end{tabular}
\end{center}
\label{tb:parameter}
\end{table}

\subsection{Semianalytic Integration}

We performed semianalytic calculations to follow the evolution of the
satellite orbits. Our procedure is the same as that used in H03. The
satellite is modeled as a single particle with variable mass and size,
and the parent as a fixed gravitational potential. The
potential of the parent is a King model with $W_0=9$ which has  the
same mass and scale as that  used in the $N$-body simulation.

For the dynamical friction, we used 
the standard 
dynamical friction formula of equation (\ref{eq:1}). 
We adopted the following 
form proposed by H03
\begin{eqnarray}
\log \Lambda = \log \left( \frac{R_{\rm {s}}}
{1.4\epsilon_{\rm s}} \right),
\label{eq:log}
\end{eqnarray}
for the Coulomb logarithm, 
where $R_{\rm {s}}$ is the distance between the center of the
parent galaxy and the satellite galaxy and $\epsilon_{\rm s}$ is the
virial radius of the satellite.

In equations (\ref{eq:1}) and (\ref{eq:log}), we use the self-bound mass
and virial radius of the satellite galaxy as $M_{\rm s}$ and
$\epsilon_{\rm s}$. For these quantities, we used the values obtained in
$N$-body simulations.

\section{Simulation Result}

\begin{figure}[htbp]
  \begin{center}  
    \FigureFile(160mm,160mm){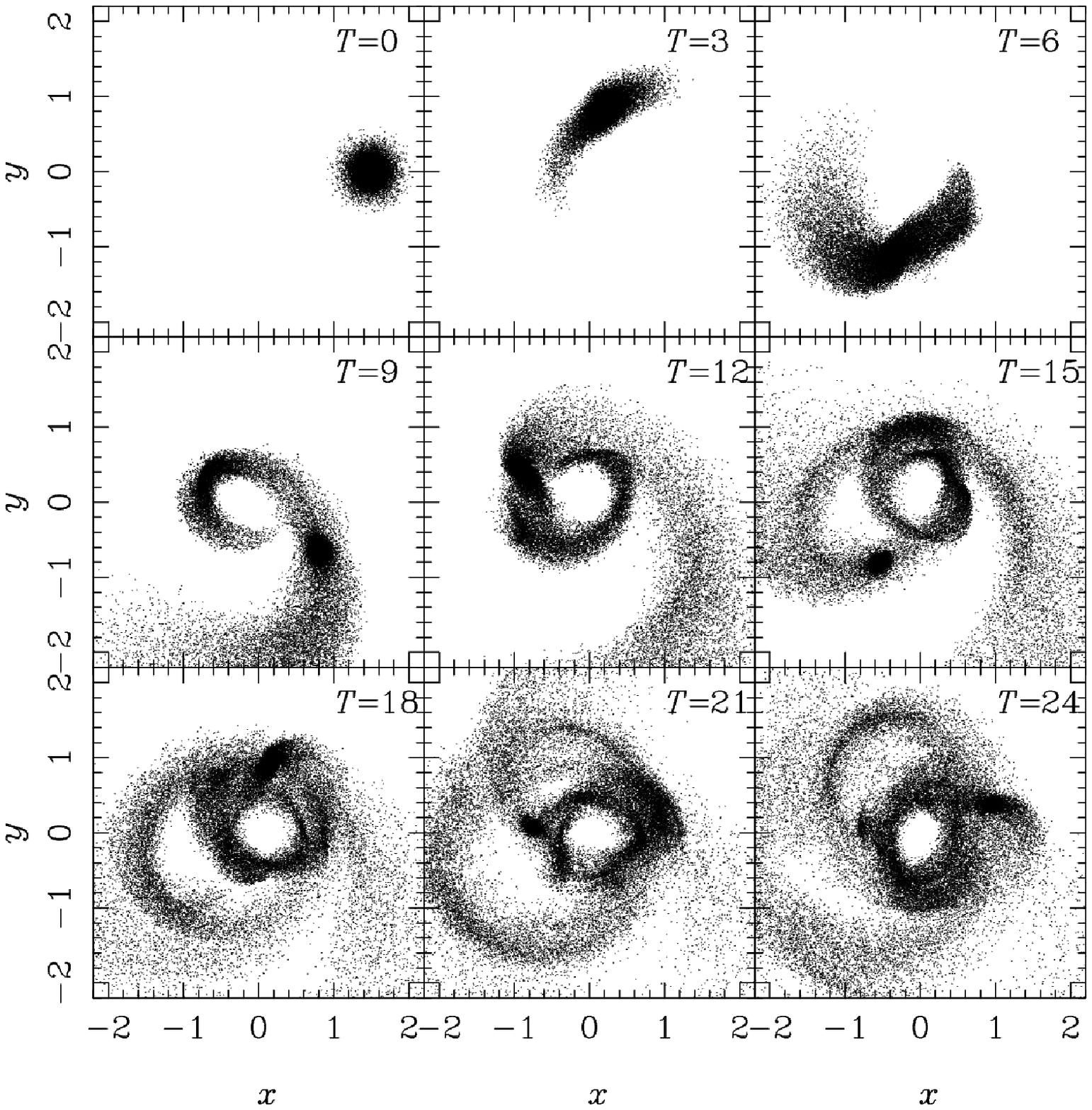}
  \end{center}
  \caption{Snapshots of the satellite particles projected onto the
 xy-plane.}
  \label{fig:snapshots}
\end{figure}

Figure \ref{fig:snapshots} shows nine snapshots of the satellite galaxy
projected  onto the x-y plane. Initially, the satellite is located at
distance 1.5 from the center of the parent galaxy. At $T=3$, it is
close to the first pericenter passage. Due to the strong tidal field
of the parent galaxy, the satellite becomes elongated. At $T=9$,
particles stripped inward and outward form clear tidal arms, and
the leading arm starts to form a circular ring. As time proceeds, more
and more mass is stripped out and at the same time the orbit of the
satellite shrinks. At $T=24$, stripped particles form complex
collection of rings and spiral patterns.

\begin{figure}[htbp]
  \begin{center}  
    \FigureFile(80mm,53mm){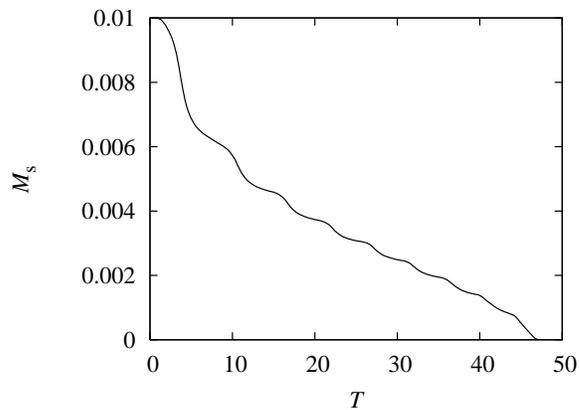}
  \end{center}
  \caption{Bound mass of the satellite $M_{\rm s}$ plotted  as a
 function of time $T$.}
  \label{fig:massloss}
\end{figure}

Figure \ref{fig:massloss} shows the evolution of the bound mass of the
satellite. At each pericenter passage, a significant amount of mass is
lost. After the pericenter passage at around $T=45$, the satellite is
disrupted.

\begin{figure}[htbp]
  \begin{center}  
    \FigureFile(80mm,53mm){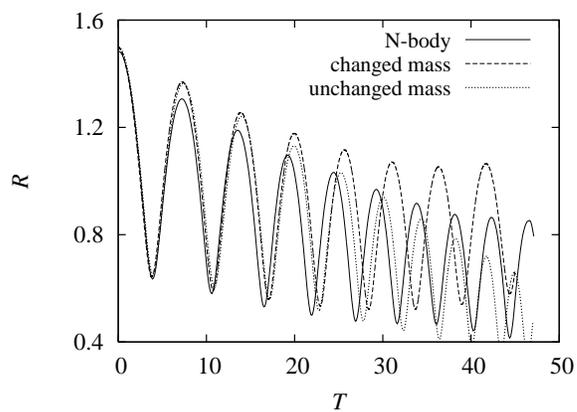}
  \end{center}
  \caption{The distance of the satellite from
  the center of the parent galaxy plotted as a function of time. 
 Solid curve shows the result of the $N$-body simulation. Dashed and
 dotted curves show that of semianalytic calculations, in which the
 mass and size of the satellite is changed ({\it dashed curve}) and 
 unchanged ({\it dotted curve}).}
  \label{fig:result1}
\end{figure}

Figure \ref{fig:result1} shows the orbital evolution of the
satellite obtained by the $N$-body simulation. We also showed the
result of semianalytic calculations. The dashed curve shows the
semianalytic calculation in which the mass and the size of the
satellite were changed using the result of $N$-body simulation. The
dotted curve showed the semianalytic calculation in which the mass
and size of the satellite were kept unchanged from their initial values.

Even though we adopted the prescription for Coulomb logarithm proposed
by H03, the agreement between the $N$-body simulation result and the
result of semianalytic orbit integration (dashed curve) is rather
poor. After the first pericenter passage, the decrease of the
apocenter distance is smaller by about a factor of two for the
semianalytic integration. This factor-of-two difference continues to
exist for entire simulation period. In fact, the $N$-body simulation
result is closer to the other semianalytic curve, for which  we ignored
the change of the mass (and the size) of the satellite, at least for
the first several orbits. Thus, taking into account the change of
the mass of the satellite somehow makes the agreement between the
$N$-body simulation and semianalytic calculation worse. Dynamical
friction formula, based on the instantaneous mass and size of the
satellite, significantly underestimates the actual drag force on the
satellite. 

This result is quite different from the results of previous
studies.  \citet{JB00} performed similar comparison between an
$N$-body simulation and a semianalytic calculation, and their result
was that the semianalytic calculation resulted in faster orbital
evolution. They used constant $\log
\Lambda$ and this must be the cause of the difference. We used
distance-dependent $\log
\Lambda$ of H03, and we found that the result is over-corrected. The
semianalytic model resulted in the orbital evolution much slower than
the result of the $N$-body simulation.

We have performed many simulations
with different initial orbits and initial satellite model, but for all
cases the result is similar. When mass loss from the satellite is
significant, the semianalytic model of H03 failed to reproduce the
orbit.

\section{Interaction between escaped particles and the satellite}

Since the difference between the H03 model and our $N$-body
simulation is that we used self-consistent model for the satellite, the
cause of the discrepancy must be the interaction between the orbital
motion of the satellite and its internal degree of freedom. There are several
ways through which the internal degree of freedom of the satellite
effectively operate as the drag force to its orbital motion. 
For example, a satellite is dynamically heated by ``bulge shock''
(Spitzer 1987) at
each pericenter passage. The energy used to heat the internal motion
of the satellite must have come from the orbital motion.

However, the internal energy of the satellite is much smaller than the
orbital energy  and not
enough to explain the orbital evolution. In the following, we consider
two mechanisms  which are potentially more efficient than simple
heating of internal motion.

The first mechanism is the interaction between the escaped particles and the
satellite. In figure \ref{fig:snapshots}, particles escaped outward
form rather impressive trailing spiral arms, while particles escaped
inward form a ring-like structure. This means the gravitational
interaction between the escaped particles and the main body of the
satellite is not symmetric. To the trailing spiral arm, the satellite
exerts some tidal torque, since the angular velocity of the satellite
is faster than that of the arm. On the other hand, the ring would not
exert much torque to the satellite, since it is axisymmetric. This
mechanism is essentially the same as the effect of non-conserving mass
transfer from a binary of two stars. The gas escaped from the $L_2$ 
point acquires the angular momentum through the interaction with the
orbital motion of the binary, resulting in the loss of  the orbital
angular momentum of the binary. In this paper we call this effect the
direct interaction between the escaped stars and the satellite.

The second one is what we named ``indirect interaction''. Many of the
particles which are stripped out of the satellite remain close to the
satellite. This is part of the reason why the direct interaction can be
important. If escapers quickly go away from the satellite, the loss of
the energy and angular momentum due to the tidal torque would be small.

If some of the escapers remain close to the satellite, they might
result in the enhancement of dynamical friction. One way to
understand dynamical friction is to regard it as the gravitational
pull by the wake of particles generated  by the satellite galaxy. The
strength of the wake depends on the mass which generates the wake. If
some escaped particles remain close to the satellite, they help making 
the wake, resulting in the enhancement of the dynamical friction.

In the following two sections, we evaluate quantitatively these two
effects in turn.

\subsection{Direct interaction with escapers}

Here, we measure the effect of the direct interaction. The
acceleration (or deceleration) of the satellite by the interaction
with the escaped particle is defined simply as
\begin{equation}
\mbox{\boldmath $a$}_{\rm di}= \frac{1}{M_{\rm s}}\sum^{i}\sum^{j}
\mbox{\boldmath $f$}_{ij},
\label{eq:direct}
\end{equation}
where the summation for $i$ is taken for particles which escaped from
the satellite, and summation over $j$ is for particles which are bound
to the satellite. The force $\mbox{\boldmath $f$}_{ij}$ is  the
gravitational force from particle $i$ to particle $j$. 

We calculate change in the specific orbital energy by this direct
interaction as
\begin{equation}
\Delta E_{\rm di} = \int_0^T\mbox{\boldmath $a$}_{\rm di}\cdot
\mbox{\boldmath $V$} dt,
\label{eq:directchange}
\end{equation}
where {\boldmath $V$} is the center-of-mass velocity of the satellite. Both
$\mbox{\boldmath $a$}_{\rm di}$ and $\mbox{\boldmath $V$}$ are
calculated from simulation result.

\begin{figure}[htbp]
\begin{center}
\FigureFile(150mm,100mm){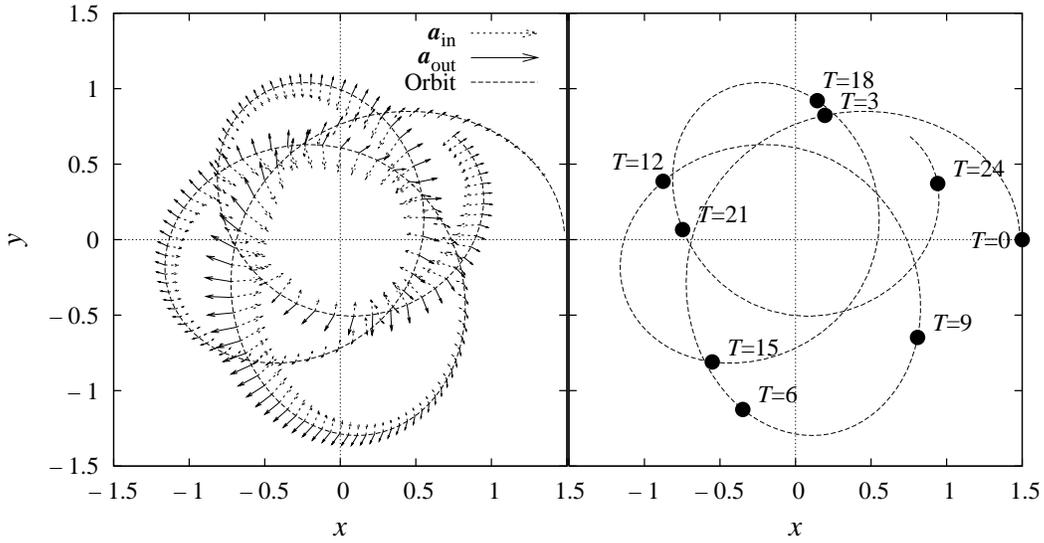}
\end{center}
\caption{The direction and strength of the direct
  effect $\mbox{\boldmath $a$}_{\rm di}$, along with the orbit of the
  satellite. The left panel shows contributions from particles outside
  the radius of the satellite and that from inside as $\mbox{\boldmath
  $a$}_{\rm out}$ and $\mbox{\boldmath $a$}_{\rm in}$ (The sum of
  these two terms gives $\mbox{\boldmath $a$}_{\rm di}$). The length
  of the arrow is 20 times the absolute value of the
  acceleration. The right panel shows the orbit of the satellite, with
  the time shown for filled circles. }
\label{fig:acc_di}
\end{figure}

Figure \ref{fig:acc_di} shows the direction and strength of
$\mbox{\boldmath $a$}_{\rm di}$ along with the orbit of the
satellite. For this figure, we separate $\mbox{\boldmath $a$}_{\rm
di}$ into the contribution of escaped particles with the distance from
the center of the galaxy larger than that of satellite (outward
escapers, $\mbox{\boldmath $a$}_{\rm out}$) and the rest (inward
escapers, $\mbox{\boldmath $a$}_{\rm in}$). By definition, 
$\mbox{\boldmath $a$}_{\rm out}$ points outwards and 
$\mbox{\boldmath $a$}_{\rm in}$ inwards. If we compare the direction
of these two terms and the orbit of satellite, we can see that the
gravitational force from outward escapers generally acts as the drag
force, while that from inward escapers changes the direction rather
often. For example, around the time of the first pericenter passage
(after $T=3$), the inward term clearly points to the direction of motion,
but it quickly changes the direction and works as the drag, until the
satellite reaches the apocenter. The outward term generally works as
the drag force, even when the satellite is going outward. 

\begin{figure}[htbp]
  \begin{center}  
    \FigureFile(80mm,53mm){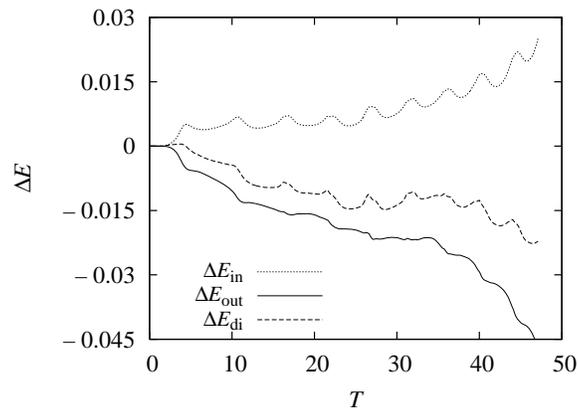}
  \end{center}
  \caption{The change of the orbital binding energy due to the direct
effect of escaped particles. Dotted, solid and dashed curves indicate
the effect of particles inside, particles outside and total effect, respectively.}
  \label{fig:edirect}
\end{figure}

Figure \ref{fig:edirect} shows the change of the orbital binding
energy of the satellite due to these forces from escaped particles. We
can see that the contribution of outward escapers works as the drag,
and that from inward escapers have the opposite effect. The total
effect is the drag. For the outward contribution, the force
and resulting energy change comes mainly from the particles which form
"trailing arm", and that is the reason why it acts as the drag force
for most of time. 

\begin{figure}[htbp]
  \begin{center}  
    \FigureFile(80mm,53mm){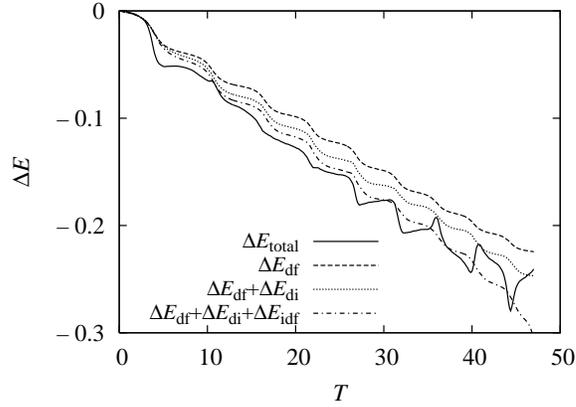}
  \end{center}
  \caption{The evolution  of the orbital energy of the satellite plotted
 as the function of time. Solid curve shows the change of the orbital
 energy $\Delta E_{\rm total}$ obtained by the $N$-body
 simulation. Dashed curve shows the estimated energy change 
 $\Delta E_{\rm df}$ due to the dynamical friction. Dotted curve
 shows the energy change due to the dynamical friction plus the direct
 effect of escaped particles $\Delta E_{\rm df} + \Delta E_{\rm di}$.
 Dash-dotted curve shows that due to the dynamical friction and the
 direct and indirect forces from escaped particles
 $\Delta E_{\rm df} + \Delta E_{\rm di} + \Delta E_{\rm idf}$.}
  \label{fig:energy}
\end{figure}

Figure \ref{fig:energy} shows the time change of $\Delta E_{\rm di}$.
For comparison, we also show the energy change due to dynamical
friction $\Delta E_{\rm df}$ calculated using equation (\ref{eq:1})
and the specific total energy change $\Delta E_{\rm total}$ obtained
from $N$-body simulation.Here, $\Delta E_{\rm total}$ is defined as
\begin{eqnarray}
\Delta E_{\rm total}=\int_0^T (\mbox{\boldmath $a$}_{\rm s}-
\mbox{\boldmath $a$}_{\rm p})\cdot \mbox{\boldmath $V$} dt, 
\label{eq:totalchange}
\end{eqnarray}
where $\mbox{\boldmath $a$}_{\rm s}$ is the center-of-mass
acceleration of the satellite, $\mbox{\boldmath $a$}_{\rm p}$ is the
acceleration due to the potential of the parent galaxy, which is estimated as
\begin{eqnarray}
  \mbox{\boldmath $a$}_{\rm p}=-\frac{GM(r)}{r^3}\mbox{\boldmath $r$},
\end{eqnarray}
where $r$ is the distance from the center of mass of the parent to
that of the satellite and $M(r)$ is the mass of the parent within $r$.
We calculated $M(r)$ from the density profile of the King model with
$W_0=9$, which is the initial  model of the parent halo in our $N$-body
simulations.

To calculate $\Delta E_{\rm total}$, we assumed that the
distribution of stars within the parent galaxy is unchanged, even
though the satellite transfers part of its  mass and orbital
energy. Since the mass of the satellite is a small fraction of the
total mass of the parent galaxy, the change of the
structure of the parent galaxy is small, and our treatment should
give fairly accurate estimate of the  energy change of the satellite. 

From figure \ref{fig:energy} we can see that the dynamical friction
of H03 formula accounts for about 80\% of the total deceleration, and
that the direct interaction accounts for about half of the remaining
20\% of the deceleration. Thus, we can conclude that the direct interaction is
significant, but other effects are not negligible. In the next section
we analyze the indirect contribution of escaped particles.

\subsection{Enhancement of the dynamical friction}

Particles which have been stripped out of the satellite but still are
close to it can increase the dynamical friction to the satellite. As
far as these ``escaped'' particles move together with the satellite,
the center-of-mass motion of these escaped particles and the satellite
feels the dynamical friction, and the strength of the dynamical
friction is determined by the total mass including the escaped
particles. A practical problem with this view is that it is not easy
to quantitatively evaluate the strength of the dynamical friction on
the bound part of the satellite galaxy. If the mass distribution of
the satellite and the escaped particles at a given time is known, we
can calculate the dynamical friction on the center-of-mass motion of
them by evaluating the linear momentum change of field
particles. However, to calculate the drag force on the bound part is a
bit complicated, since we need to evaluate the drag force from the
perturbed distribution of field stars back to the satellite.

\begin{figure}
  \begin{center}  
    \FigureFile(60mm,60mm){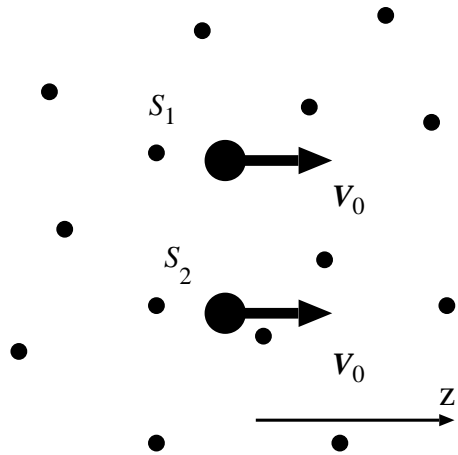}
  \end{center}
  \caption{Two massive objects moving in a uniform
  distribution of background field  particles.}
  \label{fig:config1}
\end{figure}

In the following, we give a simple model to calculate this
effect. First,  let us consider the simplest case, in which
two softened point-mass objects move in a uniform distribution
of field particles. Figure \ref{fig:config1} shows the
configuration we consider. Two massive objects, $S_1$ and $S_2$, with
equal masses 
$M$, move along $z$ axis with velocity $V_0$. 
To simplify the calculation, let us consider the case that field
particles are at rest and distributed uniformly and isotropically.
We calculate the enhancement of dynamical friction acting on $S_1$ as
a function of the distance between $S_1$ and $S_2$ in the following 
way.

The acceleration of $S_1$ due to dynamical friction is given by
\begin{eqnarray}
\frac{d{\mbox{\boldmath $v$}_{1}}}{dt} = n{\mbox{\boldmath $V$}_{0}}
\int_0^{b_{\rm max}}\!\!\! \int_0^{2\pi}
\Delta {\mbox{\boldmath $V$}_{1\parallel}}\ b_1\ d\theta_1\ db_1,
\label{eqn:dfmulti}
\end{eqnarray}
where $n$ is the number density of the field particles, 
$\mbox{\boldmath $V$}_{0}$ is the initial velocity vector of $S_1$, 
$\Delta {\mbox{\boldmath $V$}_{1}}$ is the change in velocity of $S_1$ 
caused by one encounter with a background particle,
$\Delta {\mbox{\boldmath $V$}_{1\parallel}}$ is the component of 
$\Delta {\mbox{\boldmath $V$}_{1}}$ parallel to 
$\mbox{\boldmath $V$}_{0}$, $b_1$ is the impact parameter, 
and $b_{\rm max}$ is the largest impact parameter.
(Hereafter $\parallel$ and $\perp$ mean the components parallel and
perpendicular to $\mbox{\boldmath $V$}_{0}$, respectively.)
Note that we use $b$ and $\theta$ as integration variables, which
means we chose a circle with the center at the center of coordinate
in figure 6 as the region over which we integrate the encounters. By
doing so, we made the integration region symmetric for two bodies.

\begin{figure}
  \begin{center}  
    \FigureFile(58mm,50mm){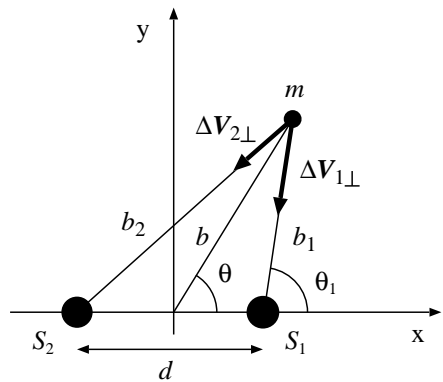}
  \end{center}
  \caption{View of two massive objects and one background particle
on the plane perpendicular to the direction of the motion of massive
particles.} 
  \label{fig:config2}
\end{figure}

For one encounter, from the momentum conservation, we have
\begin{eqnarray}
M\Delta {\mbox{\boldmath $V$}_{1\parallel}}+
M\Delta {\mbox{\boldmath $V$}_{2\parallel}}+
m \Delta {\mbox{\boldmath $V$}_{m\parallel}}=0.
\label{eq:momentum}
\end{eqnarray}
Here, $m$ is the mass of a background particle and $\Delta
 {\mbox{\boldmath $V$}_{m}}$ is its velocity change. 
Figure \ref{fig:config2} shows the view of the two massive particles
and one background particle, on the plane perpendicular to the
direction of the motion of massive particles.
Since the configuration is symmetric for two massive particles, the
dynamical friction on two particles, after integration in equation
(\ref{eqn:dfmulti}) is performed, must be equal. Thus, using
equation (\ref{eq:momentum}), we can replace
$\Delta {\mbox{\boldmath $V$}_{1}}$ in the right-hand side of
equation (\ref{eqn:dfmulti}) as
\begin{eqnarray}
\Delta {\mbox{\boldmath $V$}_{1\parallel}}=
-\frac{m}{2M} \Delta {\mbox{\boldmath $V$}}_{m\parallel}.
\end{eqnarray}

In the following, we derive the formula for $\Delta {\mbox{\boldmath
$V$}}_{m\parallel}$ using impulse approximation. It is expressed as 
\begin{equation}
  \left| \frac{\Delta {\mbox{\boldmath $V$}}_{m\parallel}}
   {{\mbox{\boldmath $V$}}_0} \right| = 1 - \cos \psi,
\end{equation}
where  $\psi$ is the deflection angle of the background particle and
is expressed as
\begin{equation}
  \psi = \left| \frac{\Delta {\mbox{\boldmath $V$}}_{m\perp}}
	  {{\mbox{\boldmath $V$}}_{0}} \right|.
\end{equation}
We can calculate $\Delta {\mbox{\boldmath $V$}}_{m\perp}$ using the
impulse approximation and by
linearly adding the contribution of two massive bodies,
$\Delta {\mbox{\boldmath $V$}}_{m1}$ and
$\Delta {\mbox{\boldmath $V$}}_{m2}$, as
\begin{eqnarray}
  \Delta {\mbox{\boldmath $V$}}_{m\perp} &=&
  \Delta {\mbox{\boldmath $V$}}_{m1\perp} +
  \Delta {\mbox{\boldmath $V$}}_{m2\perp},\\
  \Delta {\mbox{\boldmath $V$}}_{m1\perp} &=&
  \frac{2G(M+m)}{V_0}\frac{b_1}{b_1^2+\epsilon ^2}
  \hat{\mbox{\boldmath $b$}}_1,\\
  \Delta {\mbox{\boldmath $V$}}_{m2\perp} &=&
  \frac{2G(M+m)}{V_0}\frac{b_2}{b_2^2+\epsilon ^2}
  \hat{\mbox{\boldmath $b$}}_2,
\end{eqnarray}
where $\epsilon$ is the softening length and 
$\hat{\mbox{\boldmath $b$}}_1$ and
$\hat{\mbox{\boldmath $b$}}_2$ are 
the unit vectors in the directions from $m$ to $S_1$ and $S_2$ on
$xy$-plane.

\begin{figure}[htbp]
  \begin{center}  
    \FigureFile(80mm,55mm){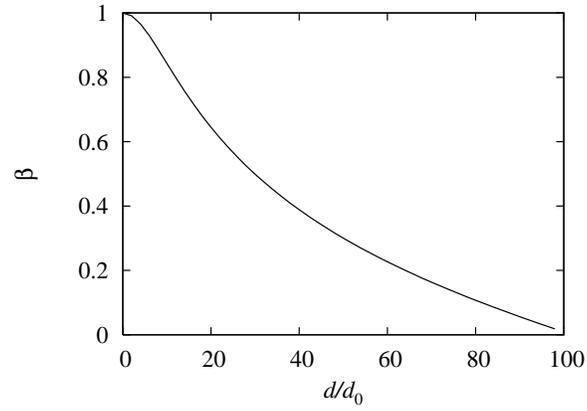}
  \end{center}
  \caption{The relative increase  of the dynamical friction $\beta$ as a
    function of the normalized distance between two objects.}
  \label{fig:friction}
\end{figure}

Using these formulae, we numerically calculated the dynamical friction
on $S_1$ as a function of the distance between two massive particles
$d$. 
Figure \ref{fig:friction} shows the enhancement $\beta$, defined as
the increase of the dynamical friction relative to the dynamical
friction of single particle moving alone, as the function of the
separation $d$. Here, 
$d_0$ is defined as follows:
\begin{eqnarray}
  d_0 \equiv \frac{MG}{V_0^2}.
\end{eqnarray}
We adopt $b_{\rm max}/d_0=100$ and $\epsilon/d_0=5.0$.
From figure \ref{fig:friction}, we can see that the increase of the
dynamical friction is significant, even when two particles are far
away (more than 10 times the softening length).

From this result, we estimated the enhancement of the dynamical
friction on the satellite due to the escaped particles. The
enhancement factor $\alpha$ is calculated as
\begin{equation}
\alpha = \frac{1}{M_{\rm s}}\int_0^{R_{\rm s}} dr\frac{dm_{\rm e}}{dr}
\beta (r),
\end{equation}
where $m_{\rm e}$ is the total mass of escaped particles within radius
$r$ from the center of mass of the satellite and $\beta (r)$ is the
value of $\beta$ at distance $r$. Note that we made many
approximations. First, we approximate the effect of particles at
distance $r$ in all directions by that of one particle in the plane
perpendicular to the direction of motion. Second, we assume the linear
relationship between the mass of the other particle and the
enhancement of the dynamical friction. Third, we assume that the
effect of multiple particles in different positions can be linearly
added.

\begin{figure}[htbp]
  \begin{center}  
    \FigureFile(80mm,55mm){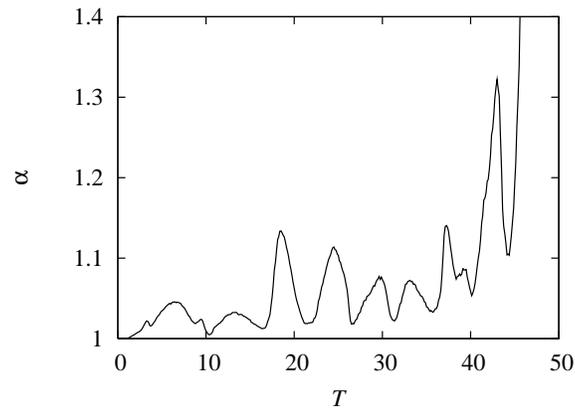}
  \end{center}
  \caption{The enhancement factor $\alpha$ as a function of time.}
  \label{fig:alpha}
\end{figure}

In figure \ref{fig:alpha} we plot the value of $ \alpha$ at each time
step in our simulation.  
The strength of enhancement changes
synchronously with the change of the distance of the satellite from the
center of the parent galaxy. The value $\alpha $ is small when the
distance is small,
{\it i.e.} near the pericenter,
while it is large when the distance is large,
{\it i.e.} near the apocenter.

The effect of this enhancement on the total energy change is shown in
figure \ref{fig:energy}. The difference between the dotted curve and the
dash-dotted one corresponds to the enhancement effect, which we call
``indirect force''. In figure \ref{fig:energy}, it is shown that the
effect of the indirect force is comparable to that of the direct force
from escapers. By taking account of the indirect effect, we can explain 
the change of the total energy quite well.

\begin{figure}[htbp]
  \begin{center}  
    \FigureFile(80mm,53mm){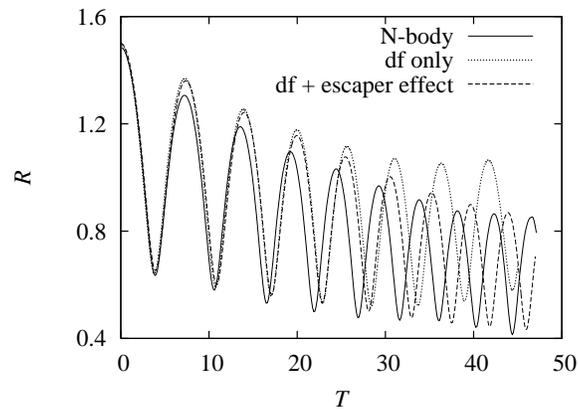}
  \end{center}
  \caption{Same as Fig. \ref{fig:result1}, but the result of
semianalytic integration using only the pure dynamical friction is
shown in the dotted curve and that with dynamical friction and both the
direct and indirect effect of escapers is shown in the dashed curve.}

  \label{fig:result2}
\end{figure}

\begin{figure}[htbp]
  \begin{center}  
    \FigureFile(80mm,53mm){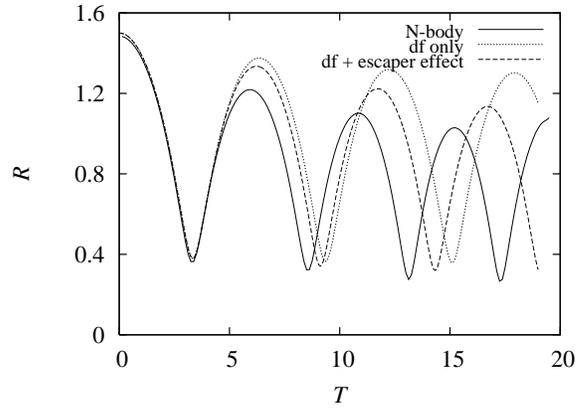}
  \end{center}
  \caption{Same as Fig. \ref{fig:result2} but the
initial velocity of the satellite is 0.326. In this calculation, the
 satellite was disrupted at around $T=20$}
  \label{fig:result3}
\end{figure}

\begin{figure}[htbp]
  \begin{center}  
    \FigureFile(80mm,53mm){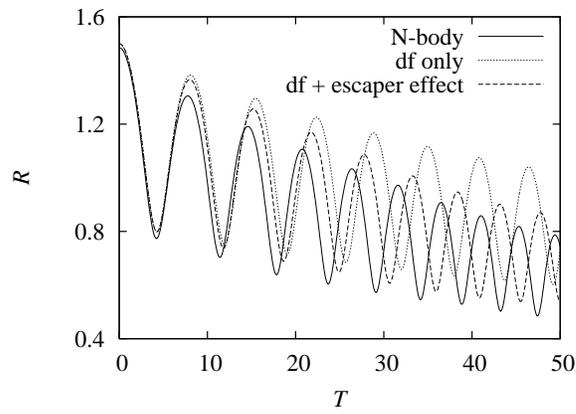}
  \end{center}
  \caption{Same as Fig. \ref{fig:result2} but the initial velocity
of the satellite is 0.5.}
  \label{fig:result4}
\end{figure}

We performed the semianalytic orbital integration using the dynamical
friction enhanced by this parameter $\alpha$. We also took into
account the direct effect of escapers which we discussed in the
previous section. 
Figure \ref{fig:result2} shows the result. The
agreement between the $N$-body simulation and the
semianalytic integration is excellent. 
Figures \ref{fig:result3} and \ref{fig:result4} show the same
comparisons for simulations from different initial orbits for the
satellite. In both cases, the agreement between  our improved
treatment and the simulation result is quite good.

\section{Summary and Discussions}

\subsection{Summary}

We studied the orbital decay of a satellite galaxy, using a fully
self-consistent $N$-body simulation in which both the satellite and
its parent galaxy are expressed by $N$-body models. We found that the
pure dynamical friction, estimated using Chandrasekhar's formula with
the correct treatment of Coulomb logarithm of the form proposed by
H03, underestimates the drag 
force by around 20\%, at least for the cases we studied.

This rather large discrepancy is due to
the effect of particles which are stripped out of the satellite by the
tidal field of the parent galaxy. They induce  additional drag
forces through two mechanisms. The first one is the direct force, which
escaped particles exert on the body of the satellite. The particles
ejected outward are accelerated by the tidal torque of the satellite,
and the satellite loses the energy and angular momentum through the
back reaction. The second mechanism is the indirect enhancement of the
dynamical friction by particles which are not bound but still in the
orbits close to the orbit of the satellite. We found these two
mechanisms have comparable contributions and the combined effect
quantitatively agrees with the discrepancy between the result of the
$N$-body simulation and the model calculation using pure dynamical
friction.

\subsection{Comparison with previous works}

\citet{JB00} compared the result of a fully self-consistent $N$-body
simulation and a semianalytic model, for the orbital decay of a satellite
galaxy. In their simulation, both the parent galaxy and the satellite were
expressed as $N$-body systems, in the same way as in our work. In the
analytic model they used, the orbital evolution was due to dynamical
friction on the bound mass of the satellite, and a simple model was
used to evaluate the mass loss due to tidal stripping.

In their work, the orbital evolution obtained with the semianalytic
model was faster than that obtained with the $N$-body simulation. This
result is the opposite to what we obtained with our first model, in
which we consider only the dynamical friction on the bound mass of the
satellite. In other words, the analytical estimate of the effect of 
dynamical friction in our model was too small, while that in
\citet{JB00} was too large.

The reason of this discrepancy is simple. When applying the
Chandrasekhar's dynamical friction formula, we used the variable Coulomb
logarithm following H03, while \citet{JB00} used the constant Coulomb
logarithm, which overestimates the dynamical friction.
\citet{Velazquez1999} also compared $N$-body simulation and model
calculation using dynamical friction formula. They obtained good
agreement, but that agreement was achieved by using the Coulomb
logarithm as a fitting parameter. H03 argued that the use of the
variable Coulomb logarithm would resolve the discrepancy between the
$N$-body simulation and the semianalytic model, without the need for
fitting parameter since Coulomb logarithm is calculated from the size of
the satellite and its distance to the center of the parent galaxy.

We found that the variable Coulomb logarithm of H03 alone would
{\it underestimate} the total drag on a live (self-consistent)
satellite, when the tidal mass loss is significant. A physically
meaningful model need to incorporate the effect of escaped
particles in some way. 

\subsection{CDM substructures}

In this paper we considered an idealized model of a spherical
satellite galaxy orbiting in a spherical parent galaxy. In the CDM
cosmology, most of the mass of a galaxy is in the CDM halo and
satellite galaxies, at least at their formation times, are in massive
CDM subhalos. Recent $N$-body simulation of the formation and
evolution of CDM halos (Kravtsov et al. 2004; Kase et al. 2006) showed
that most of subhalos lose 90\% or more of their initial mass after
they become bound to the main halo through tidal mass loss. Thus, the
mass loss they experience is typically much bigger than the mass loss
occurring to our model satellite. We can infer that the effect of mass
loss on the orbital evolution of CDM substructures or satellite
galaxies is even bigger than what we found.

\subsection{Star Clusters}

The orbits of star clusters evolve through dynamical friction. Whether
or not the effect of escaped particles are important is not
clear. Both the timescale of the orbital evolution and that of mass
loss are significantly longer than the orbital timescale. Thus, we
need more careful analysis to study these effects. In the case of very
young clusters born close to the galactic center
\citep{Fi99,McCrady2003}, we can expect the effect of the escapers to be
significant, since the ratio between the cluster mass and the relevant
mass of the parent galaxy (mass inside the location of the cluster) is
not too different from that ratio between the satellite and the parent
galaxy in our model. For most of the numerical studies of star
clusters, the pure dynamical friction formula has been used (e.g.,
Portegies Zwart and McMillan 2002; Baumgardt and Makino 2003;
G\"{u}rkan and Rasio 2005) These works might have significantly
overestimated the timescale of orbital evolution. We will study these
cases in the forthcoming paper.

\bigskip

The authors thanks Hiroyuki Kase, Keigo Nitadori and Masaki Iwasawa
for helpful discussions,  Piet
Hut, J. E. Taylor, and G. Bertin for useful comments on the manuscript,  and
Shunsuke Hozumi for detailed comments which helped us to
significantly improve the presentation of the paper. 
This
research is partially supported by the Special Coordination Fund for
Promoting Science and Technology (GRAPE-DR project), Ministry of
Education, Culture, Sports, Science and Technology, Japan.

\onecolumn

\clearpage

\begin{thebibliography}{}
  \bibitem[Barnes and Hut (1986)]{BH86}
    Barnes, J., \& Hut, P. 1986, \nat, 324, 446

  \bibitem[Baumgardt , Makino (2003)]{BM03}
    Baumgardt, H., \& Makino, J. 2003, \mnras, 340, 227

  \bibitem[Binney and Tremaine(1987)]{BT87}
    Binney, J., \&  Tremaine S. 1987, Galactic Dynamics (Princeton:
    Princeton Univ. Press), 424

  \bibitem[Bullock, Johnston (2005)]{BJ05}
    Bullock, J. S., \& Johnston, K.V. 2005, \apj, 635, 931

  \bibitem[Chandrasekhar(1943)]{C43}
    Chandrasekhar, S. 1943, \apj, 97, 255    
      
  \bibitem[Figer et al. (1999)]{Fi99}
    Figer, D. F., Kim, S. S., Morris, M., Serabyn, E., Rich, R. M., \&
    McLean, I. S. 1999, \apj, 525, 750
        
  \bibitem[Fukushige et al. (2005)]{F05}
    Fukushige, T., Makino, J., \& Kawai, A. 2005, \pasj, 57, 1009

  \bibitem[Funato et al. (1993)]{F03}
    Funato, Y., Makino, J., \& Ebisuzaki, T. 1993, \pasj, 45, 289
 
  \bibitem[G\"{u}rkan, Rasio (2005)]{GR05}
    G\"{u}rkan, M. A., \& Rasio, F. A. 2005, \apj, 628, 236

  \bibitem[Hashimoto, Funato and Makino (2003)]{H03}
    Hashimoto, Y., Funato, Y., \& Makino, J. 2003, \apj, 582, 196

  \bibitem[Heggie and Mathieu(1986)]{HM86}
    Heggie, D.C., \& Mathieu, R.D. 1986, in The Use of
    Supercomputers in Stellar Dynamics, ed. P. Hut, and S.
    McMillan (Lecture Notes in Physics 267; Berlin: Springer), 233

  \bibitem[Helmi et al. (1999)]{H99}
    Helmi, A., White, S. D. M., de Zeeuw, P. T. \& Zhao, H. 1999,
    Nature 402, 53 

  \bibitem[Ibata and Lewis (1998)]{IL98}
    Ibata, R. A., \& Lewis, G. F. 1998, \apj , 500, 575

  \bibitem[Jiang and Binney (2000)]{JB00}
    Jiang, I.-G., \& Binney, J. 2000, \mnras, 314, 468
    
  \bibitem[Johnston et al. (1995)]{KJ95}
    Johnston, K. V., Spergel, D. N., \& Hernquist, L. 1995, \apj, 451, 598
 
  \bibitem[Kase et al.(2005)]{K05}
    Kase, H., Funato, Y., \& Makino, J. 2005, in preparation
    
  \bibitem[Kravtsov et al. (2004)]{K04}
    Kravtsov, A. V., Gnedin, O. Y., Klypin, A. A. 2004, \apj, 609, 482 

  \bibitem[Makino(1991)]{M91}
    Makino, J. 1991, \pasj, 43, 621

  \bibitem[Makino (2004)]{M04}
    Makino, J. 2004, \pasj, 56, 521

\bibitem[McCrady, Gilbert and Graham (2003)]{McCrady2003}
McCrady, N., Gilbert, A.~M., \& Graham, J.~R.\ 2003, \apj, 596, 240 
 
  \bibitem[Murai, Fujimoto (1980)]{MF80}
    Murai, T., \& Fujimoto, M. 1980, \pasj, 32, 581       
   
  \bibitem[Portegies Zwart and McMillan (2002)]{PM02}
    Portegies Zwart, S. F., \& McMillan, S. L. W. 2002, \apj, 576, 899

  \bibitem[Portegies Zwart et al. (2004)]{P04}
    Portegies Zwart, S. F., Baumgardt, H., Hut, P., Makino, J., \& 
    McMillan, S. L. W, 2004, \nat, 428, 724
      
  \bibitem[Spitzer (1987)]{S87}
    Spitzer, L. 1987, Dynamical evolution of globular clusters
    (Princeton: Princeton Univ. Press), 119

\bibitem[Taylor \& Babul(2001)]{TB2001} Taylor, J.~E., \& 
Babul, A.\ 2001, \apj, 559, 716 
 


  \bibitem[van den Bosch et al. (1999)]{B99}
    van den Bosch, F. C., Lewis, G. F., Lake, G., \& Stadel, J. 1999,
    \apj, 515, 50
   \bibitem[Velazquez \& White(1999)]{Velazquez1999} Velazquez, H., \& 
White, S.~D.~M.\ 1999, \mnras, 304, 254 
 
  \bibitem[White (1976)]{W76}
    White, S. D. M. 1976, \mnras, 174, 467 

\bibitem[Zentner \& Bullock(2003)]{Z03} Zentner, A.~R., \& 
Bullock, J.~S.\ 2003, \apj, 598, 49 
 

\end{thebibliography}
\end{document}